\documentclass[twocolumn,aps,prc,showpacs,superscriptaddress,preprintnumbers,floatfix]{revtex4}
\usepackage{graphicx}
\usepackage{dcolumn}
\usepackage{bm}
\usepackage{amsmath}
\usepackage[usenames]{color}
\usepackage{ulem} 
\begin{document}
\title{Triton/$^{3}$He ratio as an observable for neutron skin thickness}
\author{ Z. T. Dai}
\affiliation{Shanghai Institute of Applied Physics, Chinese Academy of Sciences, Shanghai 201800, China}
 \affiliation{University of Chinese Academy of Sciences, Beijing 100049, China}
\author{ D. Q. Fang
}\thanks{Email: dqfang@sinap.ac.cn}
\affiliation{Shanghai Institute of Applied Physics, Chinese Academy of Sciences, Shanghai 201800, China}
\author{ Y. G. Ma
}\thanks{Email: ygma@sinap.ac.cn}
\affiliation{Shanghai Institute of Applied Physics, Chinese Academy of Sciences, Shanghai 201800, China}
\author{ X. G. Cao
}
\affiliation{Shanghai Institute of Applied Physics, Chinese Academy of Sciences, Shanghai 201800, China}
\author{ G. Q. Zhang
}
\affiliation{Shanghai Institute of Applied Physics, Chinese Academy of Sciences, Shanghai 201800, China}

\date{\today}
\begin{abstract}
Based on the framework of the Isospin-Dependent Quantum Molecular
Dynamics (IQMD) model in which the initial neutron and proton
densities are sampled according to the droplet model, the
correlation between triton-to-$^{3}$He yield ratio
(R(t/$^{3}$He)$=$Yield(t)/Yield($^{3}$He)) and neutron skin
thickness (${\delta}_{np}$) in neutron-rich projectile induced
reactions is investigated. By changing the diffuseness parameter of
neutron density distribution in the droplet model for the projectile
to obtain different ${\delta}_{np}$, the relationship between
${\delta}_{np}$ and the corresponding R(t/$^{3}$He) in
semi-peripheral collisions is obtained. The calculated results show that
R(t/$^{3}$He) has a strong linear correlation with ${\delta}_{np}$
for neutron-rich $^{50}$Ca and $^{68}$Ni nuclei. It is suggested
that R(t/$^{3}$He) could be regarded as a good experimental observable
to extract ${\delta}_{np}$ for neutron-rich nuclei because the yields
of charged particles triton and $^{3}$He can be measured quite
precisely.
\end{abstract}

\maketitle
\section{Introduction}
The proton and neutron density distributions are some of the most
fundamental properties of nuclei. Charge radii of nuclei can be
derived from charge density distributions which can be determined to
a high accuracy (often with the accuracy in charge radii better than
1\% or better for many nuclei) by experiments using electromagnetic
probes, for example, electron scattering
experiments~\cite{reference01}. The empirical information of proton
radii is then obtained from these charge radii. In contrast, our
knowledge of neutron distributions, which have been studied mainly
by hadron-nucleus scattering, is limited because the descriptions of
strong interactions in nuclei are highly model
dependent~\cite{reference02}. Reliable neutron distributions will
improve our understanding of the nucleus and nuclear
matter~\cite{reference03,reference04,reference05,reference06}.

Nuclei with neutron number ($N$) larger than proton number ($Z$) are
expected to have a neutron skin. The skin thickness is defined as
the difference between the neutron and proton root-mean-square (rms)
radii: ${\delta}_{np}=\langle r_{n}^{2}\rangle^{1/2}-\langle
r_{p}^{2}\rangle^{1/2}$. The neutron skin thickness depends on the
balance between various aspects of the nuclear force.
The formation of neutron skin arises because of the
large neutron excess and also the difference of potentials for
neutron and proton~\cite{skinform}. Strong correlations between
$\delta_{np}$ and $E_{sym}(\rho_{0})$ (nuclear symmetry energy at
saturation density $\rho_{0}$),  $L$ (the slope of symmetry energy
), $K_{sym}$( the curvature of the nuclear symmetry energy at
$\rho_{0}$), the ratio $L/J$ ($J$ is the symmetry energy coefficient
at the saturation density $\rho_{0}$), J-$a_{sym}$($a_{sym}$ is the
symmetry energy coefficient of finite nuclei) have been
demonstrated~\cite{reference03,reference07,reference08,reference09,reference10,reference11,reference12}.
These constraints are important for extrapolation of the nuclear
equation of state (EOS) to high density and hence useful for
studying properties of neutron star. Moreover, a large number of
correlations between $\delta_{np}$ and several neutron star
quantities have also been found, such as (a) the pressure of pure
neutron matter near saturation
density~\cite{reference03,reference13}, (b) neutron star
radii~\cite{reference05,reference20,reference08}, (c) the
crust-to-core transition density~\cite{reference07,reference22}, and
(d) the crustal moment of inertia~\cite{reference10,reference24}.
Neutron skin thickness is also closely related with the derivation
of volume and surface symmetry energy, as well as nuclear
incompressibility with respective to density. Furthermore, neutron
skin thickness helps to identify a nucleus with exotic structure.
Thus the precise determination of neutron skin thickness for a
nucleus becomes an important research subject in nuclear physics.

Several attempts have been made to determine neutron skin thickness.
These include hadron scattering~\cite{reference25,reference26},
$\pi^{-}$ scattering~\cite{reference27}, antiprotonic atoms
method~\cite{reference28,reference30}, giant dipole resonance (GDR)
method~\cite{reference31,reference33}, spin-dipole
resonance (SDR) method~\cite{reference34,reference35}, 
Gamow-Teller resonance (GTR) method~\cite{reference36} etc. Almost all
these methods are strongly model dependent due to the complexity of
the strong interaction between nucleons. And significant differences
exist between these experimental results. The Pb radius experiment
(PREX) at Jefferson Laboratory, has initiated a new line of research based
on the parity-violating elastic electron scattering to measure the
neutron density radius~\cite{reference37}. Although parity-violating
elastic electron scattering provides a model independent measurement
of neutron distributions, its current precision is far from
satisfactory and the method cannot be applied to unstable isotopes.

By using the Isospin-Dependent Quantum Molecular Dynamics (IQMD)
model, Sun \textit{et al}. have investigated the neutron to proton
ratio $R(n/p)$ of emitted nucleons from projectile with different
neutron skin thickness and shown that there is a strong linear
relationship between $R(n/p)$ and ${\delta}_{np}$, especially for
peripheral collisions~\cite{skinhalo1}. $R(n/p)$ is proposed as a
possible observable for extracting neutron skin size. However, it is
quite difficult to measure precisely the $n/p$ ratio experimentally
due to the low detection efficiency for neutrons. But it is
relatively much easier to measure light charged particles. Under the
coalescence picture for cluster formation, the ratio of triton to
$^{3}$He (R(t/$^{3}$He)) is expected to be proportional to the $n/p$
ratio. And R(t/$^{3}$He) can easily be measured since triton and
$^{3}$He are charged particles. Meanwhile, the ratio R(t/$^{3}$He)
is also found related with nuclear symmetry energy of neutron-rich
nuclei~\cite{symmetry1,symmetry2}. This ratio is also proposed as a
possible observable to probe the thermodynamic properties of the
fragmenting
system~\cite{temperature1,temperature2,temperature3,temperature4}.
Thus the ratio of triton to $^{3}$He is an interesting and important
physics quantity in nuclear physics. In this paper, we will explore
the relationship between the triton-to-$^{3}$He yield ratio
R(t/$^{3}$He) and the neutron skin thickness within the framework of
IQMD model. The possibility of extracting the neutron skin thickness
for neutron-rich nuclei from measurement of R(t/$^{3}$He) will be
investigated.

The paper is organized as follows. In Sec. II we briefly
describe the IQMD model, the droplet model as well as the initialization
of projectile and target. In Sec. III we present the correlation
between the neutron skin thickness and triton-to-$^{3}$He yield
ratio. Finally, a summary is given in Sec. IV.

\section{The Framework Description}
The quantum-molecular-dynamics (QMD) approach is a
many-body theory to describe heavy ion collisions from intermediate
to relativistic energies~\cite{reference38,reference103}. The main
advantage of the QMD model is that it can explicitly treat the
many-body state of the collision system and contains correlation
effects for all orders. Therefore, the QMD model can provide
valuable information about both the collision dynamics and the
fragmentation process. It mainly consists of several parts:
initialization of the projectile and the target nucleons,
propagation of nucleons in the effective potential, two body
nucleon-nucleon (NN) collisions in a nuclear medium and the Pauli
blocking.

   The IQMD model is based on the general QMD model with
explicitly inclusion of isospin degrees of freedom in the mean
field, two-body NN collisions and the Pauli blocking. In addition,
it is also important that, in initialization of the projectile and
target nuclei, the samples of neutrons and protons in phase space
should be treated separately since there exists a large difference
between neutron and proton density distributions for nuclei far from
the $\beta$-stability line. Particularly, for neutron-rich nucleus
one should sample a stable initialized nucleus with neutron-skin and
therefore one can directly explore the nuclear structure effects
through a microscopic transport model.  QMD model
has been widely and successfully used in heavy ion collisions. These
include nuclear
structure~\cite{skinhalo1,skinhalo2,skinhalo3,skinhalo4}, particle
and fragment
production~\cite{Ma1995,cluster2,fragment1,fragment2,fragment3,fragment5,fragment6},
nuclear EOS~\cite{eos1,eos2,eos3}, collective
flow~\cite{Ma-flow,flow1,flow2,flow3} and other
subjects~\cite{other1,other4,Ma-vis}.

In the IQMD model, the wave function for each nucleon is represented
by a Gaussian wave packet,
\begin{equation}
\psi _{i}(\vec{r},t_i)=\frac{2}{(2\pi
L)^{3/4}}\exp\left[-\frac{(\vec{r}-\vec{r_i}(t))^2}{4L}\right]
\exp\left[\frac{i\vec{r}\cdot\vec{p_i}(t)}{\hbar}\right],
 \label{eq1}
\end{equation}
where $\vec{r_{i}}(t)$ and $\vec{p_{i}}(t)$ are the
$i$th wave pocket in the coordinate and momentum space. $L$ is the
width of the wave pocket, which is
system-size-dependent~\cite{reference103,flow1,reference101}.
$L=2.16$ fm$^2$ is used in the present study.
And all nucleons interact via mean field and NN collisions. The nuclear
mean field can be parameterized by
\begin{eqnarray}
\lefteqn{U(\rho,\tau_z) =\alpha(\frac{\rho}{\rho_{0}})
+\beta(\frac{\rho}{\rho_{0}})^{\gamma} +\frac{1}{2}(1-\tau_z)V_c {}}
          \nonumber\\
&&\qquad\qquad {}+C_{sym}\frac{\rho_n-\rho_p}{\rho_0}\tau_z+U^{Yuk},
 \label{eq2}
\end{eqnarray}
with $\rho_{0}=0.16$ fm$^{-3}$ (the normal nuclear matter density).
$\rho$, $\rho_{n}$, and $\rho_{p}$ are the total, neutron, and
proton densities, respectively. $\tau_z$ is the $z$-th component of
the isospin degree of freedom, which equals 1 or -1 for neutrons or
protons, respectively. The coefficients $\alpha$, $\beta$ and
$\gamma$ are parameters of the nuclear EOS. $C_{sym}$ is the
symmetry energy strength due to the difference between neutron
and proton asymmetry in nuclei, which takes the value of 32 MeV.
In this paper, $\alpha =-356$ MeV, $\beta=303$ MeV and $\gamma =7/6$ are taken,
which corresponds to the so-called soft EOS. $V_c$ is the Coulomb
potential and $U^{Yuk}$ is Yukawa (surface) potential.

In the phase space initialization of the projectile and target,
the density distributions of proton and neutron are distinguished
from each other. The neutron and proton densities for the
initial projectile and target nuclei in the present IQMD model are taken
from the droplet model. In the droplet model~\cite{reference39,reference40},
we can change the diffuseness parameter to get different skin size in density distributions,
\begin{equation}
\rho_i(r)  =
\frac{\rho_{i}^{0}}{1+\exp(\frac{r-C_i}{f_{i}t_{i}/4.4})},\qquad
i=n,p,
 \label{eq3}
\end{equation}
where $\rho_{i}^{0}$ is the normalization constant which ensures
that the integration of the density distribution equals to the
number of neutrons ($i$=n) or protons ($i$=p); $t_i$ is the diffuseness
parameter; $C_i$ is the half density radius of neutron or proton
determined by the droplet model~\cite{reference40}.
\begin{equation}
C_{i} = R_{i}\left[ 1-(b_{i}/R_{i})^{2} \right], \qquad i=n,p.
\label{eq4}
\end{equation}
Here $b_i = 0.413 f_{i}t_{i}$, $R_{i}$ is the equivalent sharp
surface radius of neutron or proton. $R_{i}$ and $t_{i}$ are given
by the droplet model. In Ref.~\cite{reference30},
Trzci$\acute{n}$ska \textit{et al}. found that the half density
radii for neutrons and protons in heavy nuclei are almost the same,
but the diffuseness parameter for neutron is larger than that for
the proton which determines the neutron skin thickness.  Especially
for neutron-rich nuclei far from the stability line, a large neutron
skin is expected. Therefore, a factor $f_{i}$ is introduced by us to
adjust the diffuseness parameter. In the calculation for
neutron-rich nucleus, $f_{p}$ = 1.0 is used in Eq.(\ref{eq3}) for
the proton density distribution, while $f_{n}$ in Eq.(\ref{eq3}) is
changed from 1.0 to 1.5. Different values of $\delta_{np}$ can be
deduced from Eq.(\ref{eq3}) with different $f_{n}$ values. Using the
density distributions given by the droplet model, we can get
 the initial coordinate of nucleons in the nucleus in terms of the Monte
Carlo sampling method. In the IQMD model, the nucleon radial density
can be written as:
\begin{eqnarray}
\lefteqn{\rho(r)=\sum_{i}\frac{1}{(2\pi L)^{3/2}}\exp
(-\frac{r^{2}+r_{i}^{2}}{2L})\frac{L}{2rr_{i}} {} }
    \nonumber\\
&& \qquad {}\times \left[\exp(\frac{rr_{i}}{L})
-\exp(-\frac{rr_{i}}{L})\right],
 \label{eq6}
\end{eqnarray}
with the summation over all nucleons. And the momentum distribution
of nucleons is generated by means of the local Fermi gas approximation.
The Fermi momentum is calculated by
\begin{equation}
P_F^{i}(\vec{\mathbf{r}}) = \hbar \left[
3\pi^{3}\rho_{i}(\vec{\mathbf{r}}) \right]^{1/3} \qquad i=n,p.
\label{eq5}
\end{equation}

 To avoid taking an unstable initialization of projectile
and target in the IQMD calculation, the stability of the sampled
nuclei is strictly checked by the time evolution in the mean field
until 200 fm/c at zero temperature according to the average binding
energies, rms radii, and density distributions of the neutrons and
protons. In the initialization, the projectile and
target nucleus are treated differently. Eligible initialization
samples of projectile should meet the following requirements until
200 fm/c: ($i$) The average binding energy matches with the
experimental data; ($ii$) the rms radius is in accordance with the
droplet model; ($iii$) the neutron skin thickness of projectile
averaged on time is consistent with the droplet model.
While the target $^{12}$C only have to keep stable
in the evolution time. Using the selected initialization phase
space of nuclei in IQMD to simulate the collisions, the nuclear
fragments are constructed by a coalescence model,
in which nucleons with relative momentum smaller
than P$_0$ = 300 MeV/c and relative distance smaller than R$_0$ =
3.5 fm will be combined into a cluster. And there
are many different methods of
clusterization~\cite{cluster1,cluster2,symmetry1,Kumar2013}. Different
clustering methods may change the production rate of fragments, but
the ratio R(t/$^3$He) is less model-dependent and also less affected
by other effects~\cite{symmetry1}.
\begin{figure}[htbp]
\resizebox{7cm}{!}{\includegraphics{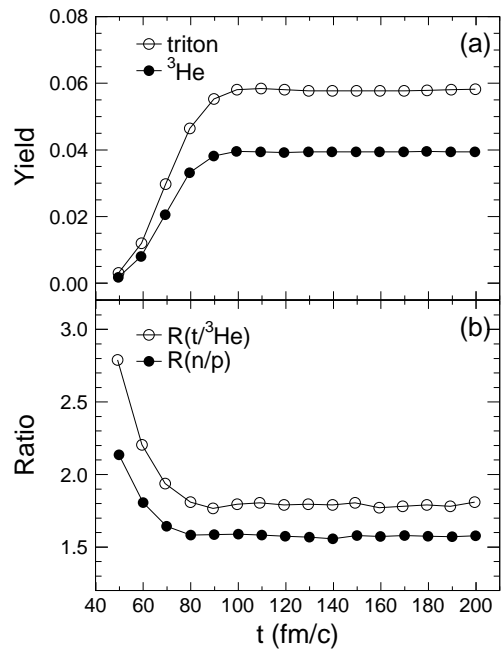}}
\caption{(a) Time evolution of the yield (per event) for triton
and $^{3}$He; (b) time evolution of R(n/p) and R(t/$^{3}$He) for
$^{68}$Ni+$^{12}$C at 50 MeV/nucleon under the condition of reduced
impact parameter from 0.6 to 1.0 and $y>0$. } \label{th_t}
\end{figure}
\begin{figure}[htbp]
\resizebox{7cm}{!}{\includegraphics{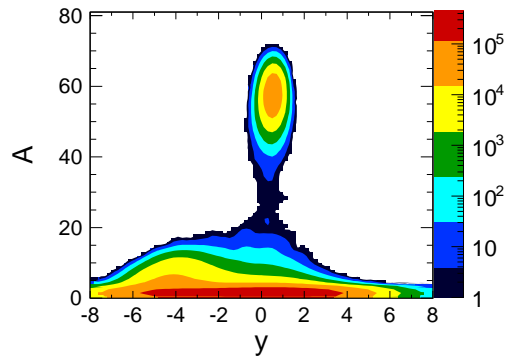}}
\caption{(Color online) Mass number versus normalized rapidity of the produced
fragments for $^{68}$Ni+$^{12}$C at 50 MeV/nucleon under the
condition of reduced impact parameter from 0.6 to 1.0 and the
evolution time from 150 fm/c to 200 fm/c.} \label{mass_y}
\end{figure}
\begin{figure}[htbp]
\resizebox{7cm}{!}{\includegraphics{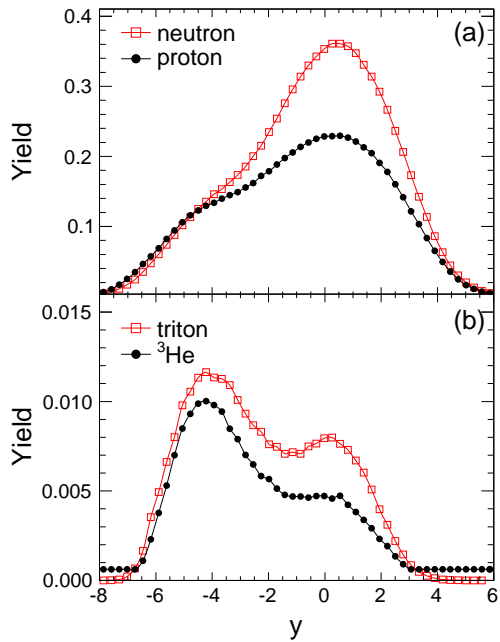}}
\caption{(Color online) (a) The normalized rapidity distribution of neutron and
proton; (b) the normalized rapidity distribution of triton and
$^{3}$He for $^{68}$Ni+$^{12}$C at 50 MeV/nucleon.
The calculation condition is the same as Fig.~\ref{mass_y}. } \label{ydis}
\end{figure}
\begin{figure}[htbp]
\resizebox{7cm}{!}{\includegraphics{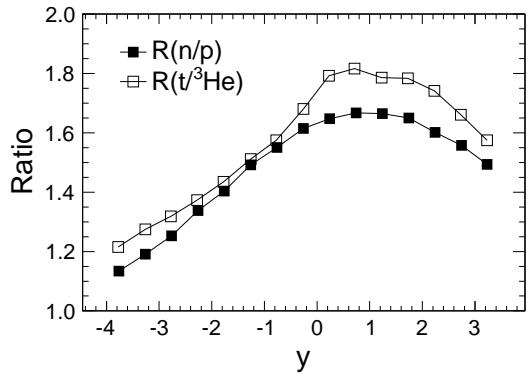}}
\caption{R(n/p) and R(t/$^{3}$He) as a function of normalized
rapidity $y$ for $^{68}$Ni+$^{12}$C at 50 MeV/nucleon.
The calculation condition is the same as Fig.~\ref{mass_y}. } \label{rnpth_y}
\end{figure}

\section{Results and Discussions}
The semi-peripheral collision processes of $^{50}$Ca and $^{68}$Ni
with $^{12}$C target at 50 MeV/nucleon are simulated using the IQMD
model. The relationship between triton-to-$^3$He yield ratio and the
neutron skin thickness in the projectile is investigated. The
fragments including neutrons and protons that formed during the
evolution of the collision are constructed by the coalescence
method. The yield ratio R(n/p) and R(t/$^3$He) can be calculated
from the emitted neutrons, protons, tritons and $^3$He. It is
assumed that the emitted fragments will have a memory of the $N/Z$
of the quasiprojectile in peripheral collisions. As the neutron skin
thickness increases, the neutron-proton composition in the surface
of nucleus will also increase. Thus the yield ratio R(t/$^3$He)
will carry the initial neutron-proton composition ($N/Z$) of its
emitting source. By changing the factor $f_{n}$ in the neutron
density distribution of the droplet model for the projectile,
different values of $\delta_{np}$ and the corresponding R(n/p) and
R(t/$^{3}$He) are obtained. Consequently, we can obtain the
correlation between R(n/p), R(t/$^{3}$He) and $\delta_{np}$.

Since the main purpose of the present work is to study the effect of
neutron skin thickness of the projectile on the yield ratio of
emitted particles. The calculation will focus on the production of
fragment from the projectile in semi-peripheral  collision. The
reduced impact parameter is used to describe the centrality of
collision which is defined as $b/b_{\text{max}}$ with
$b_{\text{max}}$ being the maximum impact parameter. 
Since the main difference between neutron and proton density
distribution is in the surface of nucleus, the
probe R(t/$^{3}$He) may be much more sensitive in peripheral
collisions. While the statistics also need to be taken into account.
Thus $0.6<b/b_{\text{max}}<1.0$ is used in the calculation. In
order to minimize the target effect on R(n/p) and R(t/$^{3}$He), we
use rapidity ($y$) cut to select neutrons, protons, tritons and
$^3$He emitted from the projectile. The rapidity of the fragments
normalized to the incident projectile rapidity is defined as:
\begin{equation}
y=\frac{1}{2} \ln \Big(\frac{E+p_{z}}{E-p_{z}} \Big)/y_{\text{proj}},
\label{eq7}
\end{equation}
where $E$ is the energy of the fragment, $p_z$ is the momentum of
fragment in $z$ direction and $y_{\text{proj}}$ is the rapidity of
the projectile. All calculations are carried out in the
center-of-mass system (CMS).

In the calculation, the time evolution of the dynamical process was
simulated until t=200 fm/c. As shown in Fig.~\ref{th_t}, the yields
of produced tritons and $^3$He (upper panel), together with the
corresponding R(t/$^{3}$He) and also R(n/p) (lower panel) are stable
after 100 fm/c. From Fig.~\ref{th_t}(b), one can see that
R(t/$^{3}$He) is larger than R(n/p) throughout the whole evolution
process. The reason may be that it is easier to combine a neutron
into a cluster than a proton due to the Coulomb repulsion
as well as the neutron-rich environment. In order
to improve statistics we accumulate the emitted neutrons, protons,
tritons and $^3$He between 150 fm/c and 200 fm/c.

While analyzing the rapidity distribution of different fragments in
semi-peripheral collisions, an interesting phenomenon  was found as
shown in Fig.~\ref{mass_y}. For a certain kind of light cluster,
there are more particles distributing in the target-like region
rather than in the projectile-like region, which is contrary to our
expectation. The rapidity distribution of neutrons, protons, tritons
and $^3$He are shown in Fig.~\ref{ydis} respectively. From the
figure one can see that there are more neutrons and protons in
projectile-like region, while for tritons and $^3$He it is
opposite. To better understand what happened in the collision
process, we followed the tracks of the nucleons in the whole
collision process. It turns out that the $^{12}$C target is much
easier to break up than the neutron-rich projectiles $^{50}$Ca
and $^{68}$Ni. This can be explained as a geometry effect. In
semi-peripheral collision of such an asymmetry system, the target is
almost penetrated by nucleons from the projectile since the size of
the projectile is much larger than the target. While only a limited
part of nucleons in the projectile is abraded.
\begin{figure}[htbp]
\resizebox{7cm}{!}{\includegraphics{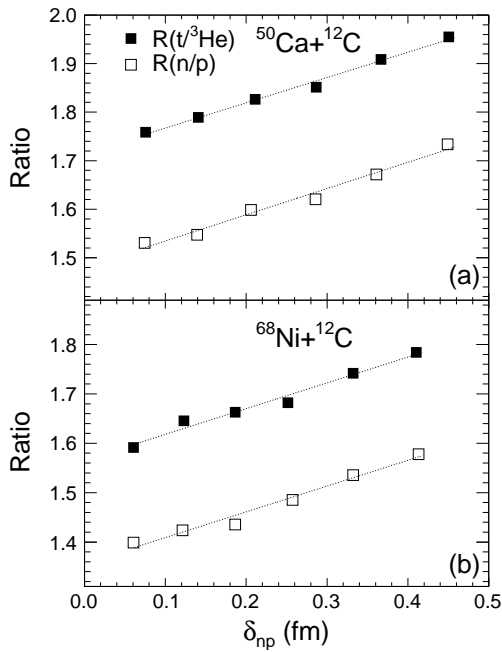}}
\caption{Dependence of R(n/p) and R(t/$^{3}$He) on neutron skin
thickness under the condition of $0.6<b/b_{\text{max}}<1.0$ and
$y>0$ for $^{50}$Ca+$^{12}$C (a) and $^{68}$Ni+$^{12}$C (b) at 50
MeV/nucleon. The dotted lines are linear fitting. } \label{ratio}
\end{figure}

\begin{figure}[htbp]
\resizebox{7cm}{!}{\includegraphics{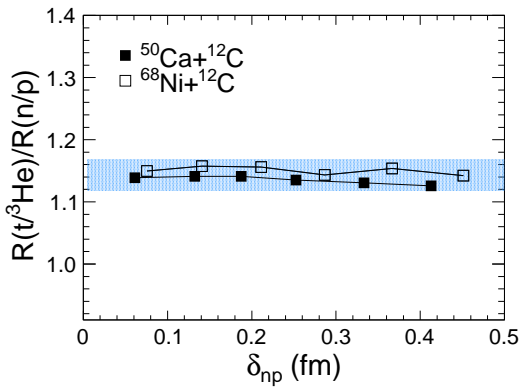}}
\caption{(Color online) Double ratio R(t/$^{3}$He)/R(n/p) as a function of
$\delta_{np}$ for $^{50}$Ca+$^{12}$C and $^{68}$Ni+$^{12}$C system
at 50 MeV/nucleon. The calculating condition is the same as
Fig.~\ref{ratio}. The blue filled area is just to guide the eyes.}
\label{rthrnp}
\end{figure}

The rapidity distribution of R(n/p) and R(t/$^{3}$He) are plotted in
Fig.~\ref{rnpth_y}. It is interesting to see that the ratio
corresponding to $A$=3 clusters indeed displays a similar rapidity
dependence to that of the emitted neutrons and protons,
though opposite rapidity dependence in their yields as discussed above.
From this figure, we can see that R(n/p) and R(t/$^{3}$He) in
projectile-like region are much larger than that in the target-like
region. This is because the neutron-rich projectile will produce
more neutrons than protons. However, what we are interested in is
the fragments produced from the neutron-rich projectile, so we make
a cut of rapidity $y>0$ to strip away fragments coming from the
target. On the other hand, the yield ratios are much more sensitive
to the surface of the nucleus, so only the events with the reduced
impact parameter from 0.6 to 1.0 are taken into account. The yields
of neutrons, protons, tritons and $^{3}$He are selected under the
condition with $0.6<b/b_{\text{max}}<1.0$ and $y>0$ and the emitted
time from 150 fm/c and 200 fm/c. The corresponding ratios
R(t/$^{3}$He) and R(n/p) with respect to the neutron skin thickness
for $^{50}$Ca+$^{12}$C and $^{68}$Ni+$^{12}$C systems are plotted in
Fig.~\ref{ratio}. A strong linear correlation between R(t/$^{3}$He),
R(n/p) and $\delta_{np}$ is exhibited, which indicates that both
R(t/$^{3}$He) and R(n/p) are sensitive to $\delta_{np}$. And
R(t/$^{3}$He) is always larger than R(n/p) as mentioned previously.

The double ratios R(t/$^{3}$He)/R(n/p) as a function of
$\delta_{np}$ for both $^{50}$Ca+$^{12}$C and $^{68}$Ni+$^{12}$C
systems are plotted in Fig.~\ref{rthrnp}. One can see that, with the
increasing of the neutron skin thickness, the double ratio
R(t/$^{3}$He)/R(n/p) is almost constant, which indicates that
R(t/$^{3}$He) is proportional to the $n/p$ ratio for different
$\delta_{np}$, which is consistent with the coalescence method. Thus
both R(t/$^{3}$He) and R(n/p) could be used as experimental
observables for determination of the neutron skin thickness.
However, R(t/$^{3}$He) will be a better quantity from the
experimental point of view since the charged particles triton and
$^{3}$He could be measured much easier than neutrons.
To see the linear relation between R(t/$^{3}$He) and the neutron skin 
thickness will be changed or not by varying the parameters in IQMD model, 
effect of the width of Gaussian wave packet ($L$) is investigated. 
The results show that a small change in $L$ has almost no effect
on R(t/$^{3}$He) and its linear dependence on neutron skin thickness.
The system of $^{48}$Ca+$^{12}$C is calculated to study the effect of 
neutrons in the projectile, the value of R(t/$^{3}$He) will decrease 
by about 7$\%$ compared with $^{50}$Ca+$^{12}$C
which is consistent with the change of $N/Z$ from $^{50}$Ca (1.5)  
to $^{48}$Ca (1.4). But the slope of linear dependence between R(t/$^{3}$He) 
and $\delta_{np}$ is the same.
Furthermore, the effect of MDI interaction is studied. With the MDI potential,
the system becomes unstable and it is very difficult to have stable 
initial samples compared with the EOS without MDI interaction.
Further investigation is necessary for understanding the effect 
of MDI interaction on the relation between R(t/$^{3}$He) and $\delta_{np}$.

\section{Summary}
Within the framework of the IQMD model in which the initial neutron and
proton densities are sampled according to the droplet model, we have
simulated the semi-peripheral collisions of 50 MeV/nucleon $^{50}$Ca
and $^{68}$Ni on a $^{12}$C target.  Assuming different neutron skin
thickness for the projectile, we have studied the correlation
between triton-to-$^{3}$He yield ratio and neutron skin thickness
for the first time. A strong linear relationship is obtained between
R(t/$^{3}$He) and $\delta_{np}$ for neutron-rich projectile, similar
with the relationship between R(n/p) and $\delta_{np}$. 
Since light charged particles could be measured easily in experiment 
as compared with neutrons, R(t/$^{3}$He) could be regarded as a sensitive 
and practical observable of $\delta_{np}$ for neutron-rich nuclei. 
If the emitted tritons and $^3$He are measured by experiment, it is
possible to extract $\delta_{np}$ from R(t/$^{3}$He). Furthermore,
some information about the nuclear equation of state could be
deduced after the determination of neutron skin thickness from
the experimental measurements.

\section*{Acknowledgements}
This work was supported in part by the Major State Basic
Research Development Program in China under Contract No. 2013CB834405,
the National Natural Science Foundation of China under Contract 
No.s 11175231 and 11035009, and the Chinese Academy of Science 
Foundation under Grant No. KJCX2-EW-N01.

\end{document}